\documentclass{SciPost}

\hypersetup{
    colorlinks,
    linkcolor={red!50!black},
    citecolor={blue!50!black},
    urlcolor={blue!80!black}
}

\usepackage[bitstream-charter]{mathdesign}
\DeclareSymbolFont{usualmathcal}{OMS}{cmsy}{m}{n}
\DeclareSymbolFontAlphabet{\mathcal}{usualmathcal}

\fancypagestyle{SPstyle}{
\fancyhf{}
\lhead{\colorbox{scipostdeepblue}{\bf \color{white} ~SciPost Physics Proceedings }}
\rhead{{\bf \color{scipostdeepblue} ~Submission }}

\fancyfoot[C]{\textbf{\thepage}}
}

\begin{document}

\pagestyle{SPstyle}

\begin{center}{\Large \textbf{\color{scipostdeepblue}{
Rings of Light, Speed of AI: YOLO for Cherenkov Reconstruction\\
}}}\end{center}

\begin{center}\textbf{
Martino Borsato\textsuperscript{1,2},
Giovanni Lagan\'a\textsuperscript{1$\star$} and
Maurizio Martinelli\textsuperscript{1,2$\dagger$}
}\end{center}

\begin{center}
{\bf 1} University of Milano-Bicocca
\\
{\bf 2} INFN, Sez. di Milano-Bicocca
\\[\baselineskip]
$\star$ \href{mailto:giovannilag66@gmail.com}{\small giovannilag66@gmail.com}\,,\quad
$\dagger$ \href{mailto:maurizio.martinelli@unimib.it}{\small maurizio.martinelli@unimib.it}
\end{center}

\definecolor{palegray}{gray}{0.95}
\begin{center}
\colorbox{palegray}{
  \begin{tabular}{rr}
  \begin{minipage}{0.37\textwidth}
    \includegraphics[width=60mm]{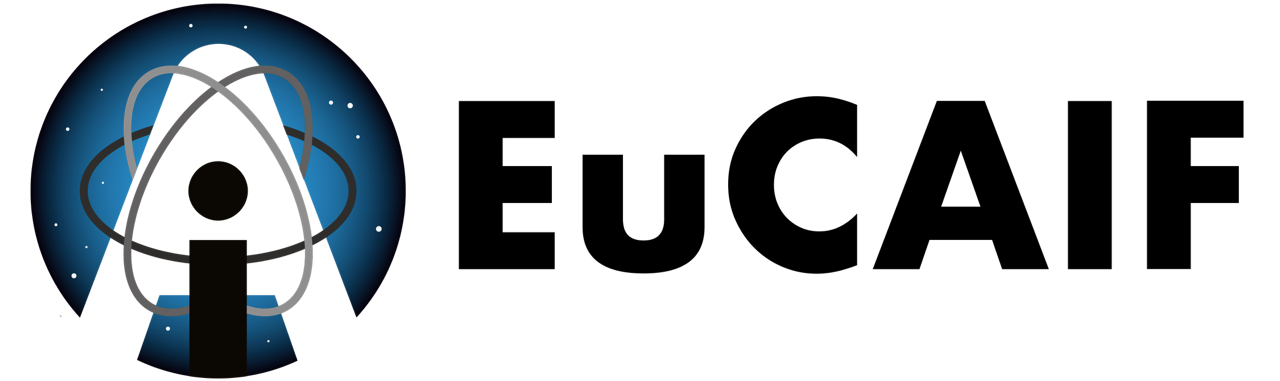}
  \end{minipage}
  &
  \begin{minipage}{0.5\textwidth}
    \vspace{5pt}
    \vspace{0.5\baselineskip} 
    \begin{center} \hspace{5pt}
    {\it The 2nd European AI for Fundamental \\Physics Conference (EuCAIFCon2025)} \\
    {\it Cagliari, Sardinia, 16-20 June 2025
    }
    \vspace{0.5\baselineskip} 
    \vspace{5pt}
    \end{center}
    
  \end{minipage}
\end{tabular}
}
\end{center}

\section*{\color{scipostdeepblue}{Abstract}}
\textbf{\boldmath{
Cherenkov rings play a crucial role in identifying charged particles in high-energy physics (HEP) experiments. 
Most Cherenkov ring pattern reconstruction algorithms currently used in HEP experiments rely on a likelihood fit to the photo-detector response, which often consumes a significant portion of the computing budget for event reconstruction. 
We present a novel approach to Cherenkov ring reconstruction using YOLO, a computer vision algorithm capable of real-time object identification with a single pass through a neural network. 
We obtain a reconstruction efficiency above 95\% and a pion misidentification rate below 5\% across a wide momentum range for all particle species.
}}

\vspace{\baselineskip}

\noindent\textcolor{white!90!black}{%
\fbox{\parbox{0.975\linewidth}{%
\textcolor{white!40!black}{\begin{tabular}{lr}%
  \begin{minipage}{0.6\textwidth}%
    {\small Copyright attribution to authors. \newline
    This work is a submission to SciPost Phys. Proc. \newline
    License information to appear upon publication. \newline
    Publication information to appear upon publication.}
  \end{minipage} & \begin{minipage}{0.4\textwidth}
    {\small Received Date \newline Accepted Date \newline Published Date}%
  \end{minipage}
\end{tabular}}
}}
}


\vspace{10pt}
\noindent\rule{\textwidth}{1pt}
\tableofcontents
\noindent\rule{\textwidth}{1pt}
\vspace{10pt}

\section{Introduction}
\label{sec:intro}
When a charged particle traverses a medium at a velocity exceeding the phase velocity of light, it emits Cherenkov radiation \cite{PhysRev.52.378} in the form of conical wakes of light. The size of the light cone depends directly on the mass and momentum of the particle that produced it. In particular, the Cherenkov angle $\theta_C$ is related to the particle’s velocity by:

\begin{equation}
\cos \left(\theta_C\right)=\frac{v_p}{v}=\frac{1}{n \beta},
\label{eq:cosine_of_cherenkov_angle}
\end{equation}
where \textit{n} is the refractive index of the medium, and $\beta=v/c$.
This relation shows that measuring $\theta_C$ provides direct access to the particle velocity, which, when combined with the momentum from tracking detectors, allows to estimate the particle's mass, and hence to identify its species.

At the LHCb experiment \cite{LHCb:2023hlw}, this principle is exploited by the Ring Imaging Cherenkov (RICH) detectors, which capture the Cherenkov photons as characteristic ring patterns on the photodetector plane. LHCb employs two such detectors (RICH1 and RICH2), optimized for complementary momentum ranges, ensuring efficient particle identification (PID) across the experiment’s acceptance. Accurate PID is indispensable for suppressing combinatorial background and enhancing event selection, and for this reason the RICH response is integrated directly into the online trigger. However, reconstructing the rings with the traditional maximum-likelihood method is computationally intensive: in the high-level trigger (HLT2), RICH PID consumes about $20{-}40\%$ of the CPU budget, making it one of the major bottlenecks for real-time operation. At Run 3 luminosities, the LHCb trigger reduces the $30$ million proton–proton collisions per second to an HLT2 input rate of $0.5{-}1.5$ million events per second \cite{LHCB-FIGURE-2020-016}. Since each event typically contains on the order of one hundred Cherenkov rings, the RICH PID task requires processing $\mathcal{O}(10^8)$ ring classifications every second.
To address those computational challenges we propose a YOLO-based pipeline trained to directly infer particle species from ring images. Similar ML-based approaches have already been explored with promising results \cite{Blago:2791645}, showing the potential of deep learning for RICH PID.

\section{Dataset}

The training and evaluation data are produced with a dedicated synthetic event generator~\cite{RICHGenerator} that emulates the response of the RICH1 detector (the procedure transfers straightforwardly to RICH2). Each simulated event consists of a set of Cherenkov rings projected onto the photodetector plane.

For every event, the generator samples a random number of primary particles (between 160 and 180) from LHCb-like momentum distributions. Each particle is assigned a species (pion, kaon, proton, electron, or muon) and a momentum. From these parameters, the Cherenkov angle is computed using Eq.~\ref{eq:cosine_of_cherenkov_angle}, and the corresponding ring radius $R$ on the detector plane is
\begin{equation}
R=\frac{\tan \left(\theta_c\right) \cdot R_{\max }}{\tan \left(\theta_{c_{\max }}\right)},
\label{eq:cherenkov_ring_radius}
\end{equation}
where $R_{\text {max }}$ is the maximum ring radius ($100 \mathrm{~mm}$ was used to roughly emulate the optics of RICH1), and $\theta_{c_{\max }}$ is the maximum Cherenkov angle ($\beta=1$).

Ring centers are then sampled from a distribution reproducing the geometry and acceptance of the RICH1 detector, ensuring a realistic spread across the sensor plane. The expected photon yield for each ring is proportional to $\sin^2(\theta_C)$ and is set to a maximum of 60 hits. Photon hits are radially smeared with a Gaussian resolution of 1.5 mm (0.8~mrad). No Cherenkov hit is generated for the ring centers of particles that have a speed below threshold.

\begin{figure}[t]
  \centering
  \includegraphics[width=0.8\linewidth]{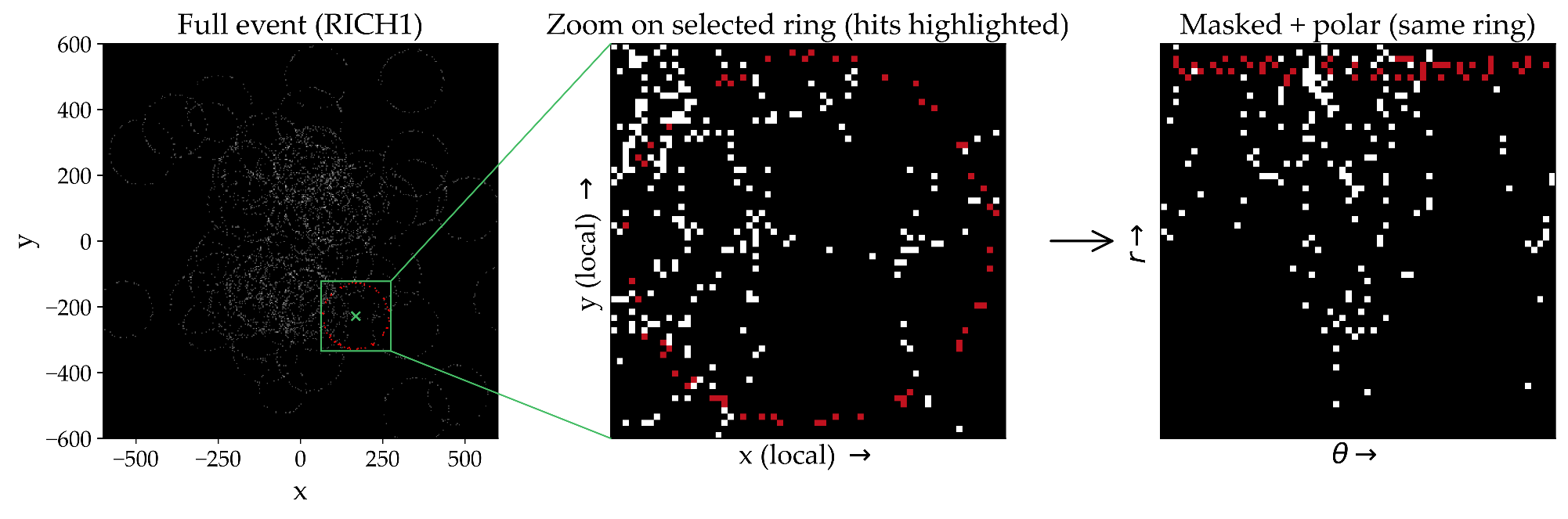}
    \caption{From full RICH1 event to masked polar view. 
    \textbf{Left}: full event (selected ring in red); 
    \textbf{middle}: single-ring image in Cartesian coordinates with highlighted hits; 
    \textbf{right}: annular mask and polar transform $(r,\theta)$. 
    In middle and right, all lit pixels are set to 255 for clarity.}
  \label{fig:event-with-ring-image}
\end{figure}

For the YOLO models, we generate images centered around each tracked ring (to emulate the LHCb RICH1 tracking efficiency, only $\sim32 \%$ of the rings are retained as tracked in the test set). Before rasterization, an annular mask is applied to restrict the image content to the radial region compatible with possible Cherenkov radii at the given momentum, thereby suppressing irrelevant background. The hits are then transformed into polar coordinates, which linearizes the circular rings into horizontal lines (see figure \ref{fig:event-with-ring-image}).

Finally, to build a balanced training dataset, generated events are used as realistic backgrounds, on top of which a controlled number of signal rings of specific momentum and particle type are added. This procedure allows us to produce class-balanced samples for each momentum bin while retaining the realism of events.

\section{YOLO Classification Pipeline}

\textit{YOLO} (You Only Look Once)~\cite{ultralytics_yolo,redmon2016lookonceunifiedrealtime} is a family of single-pass convolutional models designed for real-time vision tasks. In this work we use the \emph{classification} variant (\textit{YOLO-cls}) as a fast per-ring classifier within a momentum-binned PID scheme.

As discussed in Sec.~\ref{sec:intro}, the Cherenkov radius is a deterministic function of $\beta$ and thus of momentum at fixed mass. We partition the full momentum range into bins that fulfill the following requirements: the expected radii of distinct particle species do not overlap, and for each particle type the variation of the radius within the bin’s momentum range is below a threshold. These conditions ensure that each image corresponds to a unique particle identification, avoiding multiple possibilities. Each bin is then treated as an independent, closed-set classification problem with its own label space.
In addition, for each bin, we merge species with very close Cherenkov radius into the same class. 
We tested both strategies---training with merged classes in these ambiguous regions and training with all particle types kept separate---and compared their performance, finding that the grouping approach yields higher accuracy.

\subsection{Training and evaluation}

For each momentum bin we train an independent YOLOv11-cls network, initialized from the pretrained Ultralytics weights \cite{ultralytics_yolo}. The loss function is the standard cross-entropy over the bin’s label set. To ensure balance, the training sets are constructed with a uniform distribution of particle types and momenta, consisting of $25\,000$ images per bin split into $80\%$ training and $20\%$ validation.

All models corresponding to the different bins are loaded onto the device. During inference, each ring image is dispatched to the appropriate model according to its momentum. While effective, this dispatching procedure introduces a non-negligible overhead, as we will discuss later.

The global test set is built from $850\,000$ rings with realistic momentum and particle-type distributions. Evaluation focuses primarily on the pion misidentification rate (pion misID) versus the kaon identification efficiency (Kaon ID), the standard benchmark in RICH PID.

\subsection{Experiments}

We conducted three main sets of experiments to evaluate different design choices of the YOLO pipeline and to identify the optimal configuration for RICH PID.

\paragraph{Grouping vs non-grouping.}
As discussed earlier, at high momenta the Cherenkov radii of different particle species converge to nearly identical values, making them physically indistinguishable. We therefore tested two training strategies: 
(i) keeping each particle type as a separate class across all bins, and 
(ii) merging indistinguishable types into a single class, with probabilities later split among the hypotheses at inference time. 
The results show that class grouping significantly improves stability and reduces misidentification rates at low and high momentum. Consequently, the grouped configuration was adopted for the following experiments.

\paragraph{Model size.}
YOLOv11 comes in multiple size variants, from \textit{nano} to \textit{extra-large}, trading off accuracy against computational cost. We compared the \textit{nano}, \textit{small}, and \textit{medium} models on identical training and test samples. Surprisingly, the \textit{medium} variant did not outperform the smaller models: in the momentum range $6{-}60$ GeV the \textit{small} and \textit{nano} variants achieved lower pion misidentification rates and similar kaon efficiencies, while also providing faster inference. We attribute this to the increased complexity of the \textit{medium} model, which may require more data and training to converge optimally. Given its balance of accuracy and efficiency, the \textit{small} variant was selected as the baseline.

\paragraph{Image size.}
We also investigated the effect of input resolution, training models with images of $32\times32$, $64\times64$, and $128\times128$ pixels. Larger images capture more detail and generally improve accuracy, with the $128\times128$ model giving the lowest pion misID rates. However, this resulted in increased inference times and higher memory consumption. The $64\times64$ configuration provided a good compromise, offering nearly the same accuracy as $128\times128$ while maintaining fast inference comparable to the $32\times32$ setup.

\paragraph{Optimal pipeline.}
\label{sec:optimal_pipeline}
From these studies we conclude that the best configuration consists of the YOLOv11-small model trained on $64\times64$ images, with grouped particle classes in high-momentum bins.

\subsection{Comparison with Maximum-Likelihood fit}

\begin{figure}[ht]
  \centering
  \includegraphics[width=0.6\linewidth]{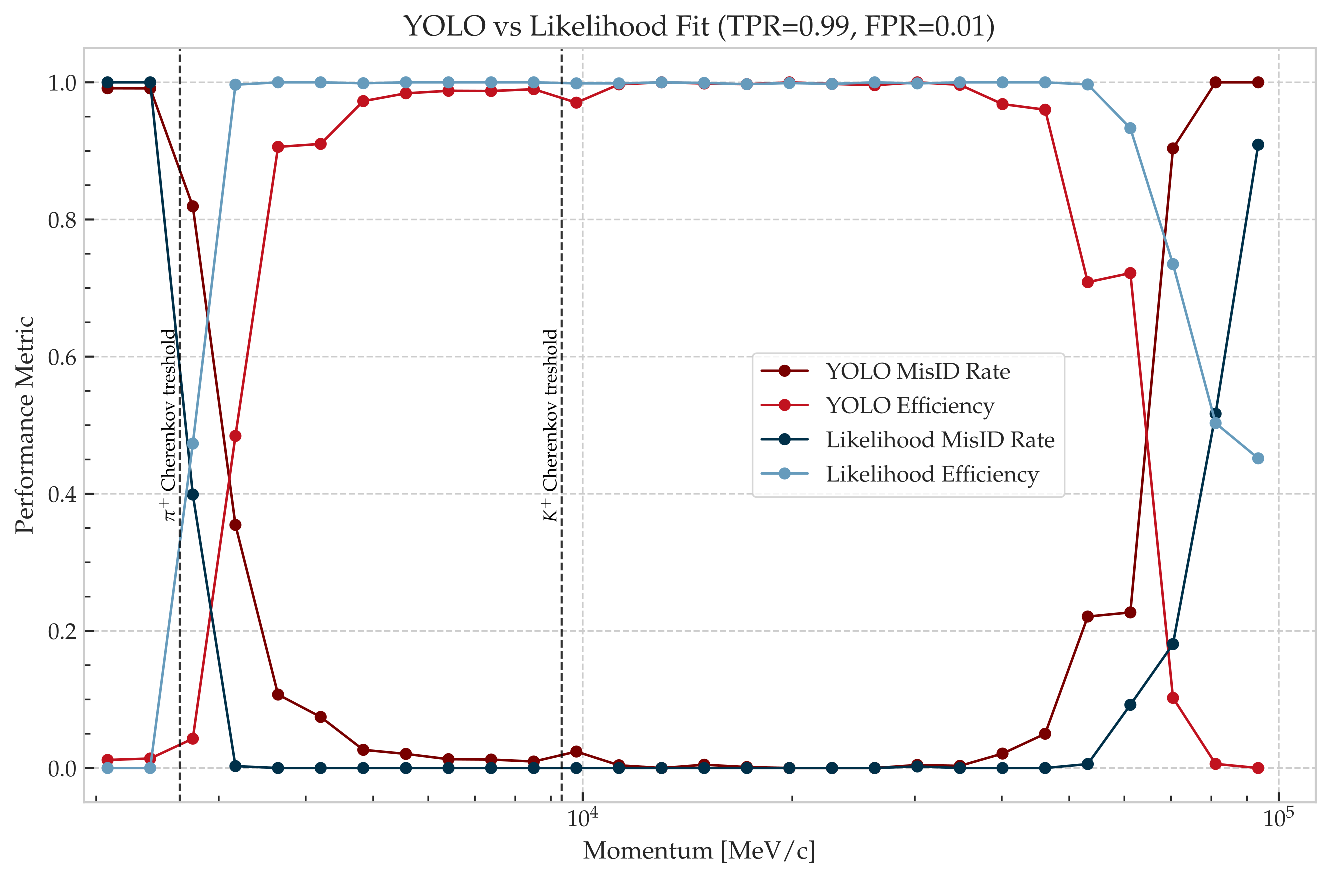}
  \caption{Comparison of YOLO Object Detection Performance and Likelihood Estimation.}
  \label{fig:yolo_vs_llh}
\end{figure}

Figure~\ref{fig:yolo_vs_llh} presents a comparison between the YOLO pipeline and the maximum-likelihood (ML) fitter at a fixed operating point (TPR$=0.99$, FPR$=0.01$). The separation of kaons and pions as a function of the track momentum is used as a benchmark. In the central momentum region, where the Cherenkov separation between species is maximal and photon yield is high, the two approaches perform almost identically: kaon identification efficiencies remain above $95\%$ while pion misID rates stay at or below the percent level. At lower momenta, close to the pion threshold, the ML fitter reaches its plateau more quickly and maintains lower misID than YOLO, which shows a broader transition region. At higher momenta, where Cherenkov radii saturate and inter-species separation decreases, ML again provides better accuracy, with YOLO showing a gradual loss of efficiency and a corresponding increase in misID. 
Overall, YOLO provides competitive performance in the momentum range most relevant for RICH1 physics. However, across the full spectrum, the likelihood-based approach achieves higher accuracy. This is because it leverages the complete hit information from the entire detector plane to infer the PID of all particles simultaneously, whereas YOLO relies only on the local subset of hits around the track under consideration.

\section{Conclusions}

We have presented a YOLO-based pipeline for Cherenkov ring reconstruction in the LHCb RICH detectors. Using the \textit{optimal} configuration (Sec.~\ref{sec:optimal_pipeline}), we achieved kaon identification efficiencies above $95\%$ at a pion misID rate of $1\%$ over a broad momentum range, with an average inference time of $\sim 20$\,ms per full event (118 rings) on an NVIDIA A6000 GPU.

These results are very promising in terms of both accuracy and speed. The main limitation we identified is the dispatching mechanism, required by the bin-wise training of models, which introduces a non-negligible overhead during inference. To quantify the dispatching overhead, we ran inference on a single bin without it, obtaining an estimated inference time per event of $\sim 4 \text{ms}$, confirming that dispatching dominates the inference time.

Next steps include modifying the YOLO model so that it takes as input, in addition to the ring image, the particle momentum and the ring center position on the detector. This will make it possible to train a single unified model without binning, thereby removing the dispatching bottleneck and enabling an even faster and more scalable RICH PID pipeline.
Furthermore, we plan to train the YOLO model on PID calibration datasets where one of the tracks is identified by fully reconstructing the decay it comes from. Since only a single track per event can be tagged, this procedure is not viable to tune the maximum-likelihood based model and could represent an advantage of the YOLO approach to PID. 
The capability of identifying the PID of a single track very quickly with YOLO by using only the photon hits around its center could be useful in order to use the PID information to refine the first trigger-level decisions.

\bibliography{SciPost_Example_BiBTeX_File.bib}

\end{document}